\documentclass{IEEEtran}
\usepackage{cite}
\usepackage{amsmath,amssymb,amsfonts}
\usepackage{algorithmic}
\usepackage{graphicx}
\usepackage{textcomp}
\def\BibTeX{{\rm B\kern-.05em{\sc i\kern-.025em b}\kern-.08em
    T\kern-.1667em\lower.7ex\hbox{E}\kern-.125emX}}
\begin{document}
\title{ Continuous and Discrete-Time Filters: A Unified Operational Perspective  }

\author{Luca Giangrande
\thanks{
L. Giangrande is with CERN , Esplanade de Particules 1, Meyrin, Switzerland (e-mail: luca.giangrande@cern.ch) 
and with the University of Geneva, Department of nuclear and particle physics DPNC, Quai Ernest Ansermet 24, 1211 Geneva (e-mail: luca.giangrande@unige.ch). }}

\maketitle

\begin{abstract}
Continuous-time (CT) and discrete-time (DT) linear time-invariant (LTI) systems are commonly introduced through distinct mathematical formalisms, which can obscure their underlying dynamical equivalence. This tutorial presents a unified treatment of first-order CT and DT systems, emphasizing their shared modal structure and stability properties. Beginning with transfer functions and pole–zero representations in the Laplace domain, canonical first-order low-pass and high-pass dynamics are examined from an operational perspective. The discussion then transitions to discrete-time sequences and the Z-transform, highlighting geometric sequences as eigenfunctions of DT systems and establishing the correspondence between the left half of the s-plane and the interior of the unit circle in the z-plane. Practical discretization and sampled-data implementations are analyzed to illustrate how continuous-time dynamics are reinterpreted through recursion and accumulation in discrete-time realizations. By maintaining structural symmetry between domains, the manuscript consolidates established concepts into a coherent framework linking mathematical representation, physical realizability, and implementation.
\end{abstract}


\section{Introduction}
Continuous-time (CT) and discrete-time (DT) signal processing are typically introduced through distinct mathematical frameworks: differential equations and Laplace transforms for CT systems, difference equations and Z-transforms for DT systems [1]–[9]. While this separation is natural from a formal standpoint, it can obscure the fact that both domains describe the same underlying linear dynamical behavior under different operational constraints.

In practice, engineers frequently move between CT and DT representations when designing filters, control loops, and mixed-signal systems. Transfer functions, poles and zeros, impulse responses, and stability criteria reappear in both settings, yet their physical interpretation often shifts subtly with sampling, delay, and implementation architecture. As a result, systems that are well understood in one domain may exhibit unexpected behavior when translated into the other.

This manuscript presents a unified tutorial treatment of CT and DT linear time-invariant (LTI) systems, with emphasis on their shared dynamical structure. The discussion is intentionally restricted to single-input single-output systems and focuses primarily on first-order dynamics. Despite their simplicity, first-order systems already capture integration, differentiation, filtering, stability, bandwidth limitation, and delay. They therefore provide a minimal yet expressive framework for examining how continuous and discrete descriptions correspond.

The analysis begins in continuous time, revisiting transfer functions, pole–zero representations, and impulse responses from an operational viewpoint. First-order low-pass and high-pass filters are examined as canonical realizations of integration and differentiation under physical constraints. The discussion then transitions to discrete-time sequences, introducing the Z-transform and discrete Fourier representations as tools for modeling sampled systems. Discrete-time LTI systems are developed in parallel with their continuous-time counterparts, highlighting the role of geometric sequences as eigenfunctions of difference equations and the stability condition imposed by the unit circle.

The connection between domains is made explicit through pole mapping between the Laplace and Z planes and through practical discretization strategies. Sampled-data filter architectures are finally illustrated using modular implementations based on delays, summations, and feedback, emphasizing how CT dynamics are reinterpreted through accumulation and recursion in DT systems.

Rather than introducing new theoretical results, the goal is to consolidate established concepts into a coherent framework that links mathematical representation, physical realizability, and implementation structure. By maintaining symmetry between the Laplace and Z domains, the paper aims to provide a unified intuition that remains valid across analog, sampled-data, and fully digital realizations.

\section{Continuous-Time Transfer Functions}
To maintain symmetry with later sections of the manuscript, details of CT modeling are recalled here for clarity.
A CT LTI system is fully described by the linear differential equation relating the input, $x_{in}(t)$,  to the output signal, $x_{out}(t)$:

\begin{equation}
    \sum_{k=0}^{n} a_k \frac{d^k x_{\text{out}}(t)}{dt^k} = \sum_{j=0}^{m} b_j \frac{d^j x_{\text{in}}(t)}{dt^j}
    \label{eq:ct-lti_diff}
\end{equation}

The constant coefficients $b_j$ and $a_k$ weight the input and output derivatives, respectively. 
While this differential representation directly encodes the physical behavior, the Laplace transform offers clearer insight into key properties such as stability, transient evolution, and frequency response. By converting differential relations into algebraic ones, it eliminates the need to handle derivatives explicitly. Applying the Laplace transform to both sides of Eq.~\ref{eq:ct-lti_diff} and grouping terms yields a polynomial ratio, in the complex variable $s$, that encodes the input-output relationship: 

\begin{equation}
    \frac{\sum_{j=0}^{m} b_j s^j }{\sum_{k=0}^{n} a_k s^k }=\frac{X_{\text{out}}(s)}{X_{\text{in}}(s)} \triangleq H(s)
\label{eq:TF_Laplace}
\end{equation}
and is therefore termed the transfer function of the system. The convolution property and linearity allow to obtain the time-domain output via the inverse transform expression:

\begin{equation}
    x_\textit{out}(t) = \mathcal{L}^{-1} \left[ H(s) \cdot X_\textit{in}(s)\right] = \mathcal{L}^{-1} \left[ H(s)\right] * \mathcal{L}^{-1} \left[X_\textit{in}(s)\right]
    \label{eq:anti_transf}
\end{equation}
where $*$ represents the convolution operation. Using the transform of Dirac's delta function, $\delta(t)$, it can be shown how $\mathcal{L}^{-1}[H(s)]$ corresponds to the time-domain output in response to an ideal impulse:

\begin{equation}
    \mathcal{L}^{-1} \left[ H(s)\cdot 1\right]= \int_{-\infty}^{+\infty} h(t-\hat{t})\,\delta(\hat{t})\cdot d\hat{t}=h(t)
    \label{eq:impuls_resp}
\end{equation}
This impulse response, $h(t)$, also encodes a complete description of the system and can be used to compute the output for an arbitrary input entirely in the time-domain via the convolution integral:

\begin{equation}
    x_\textit{out}(t) = h(t)* x_\textit{in}(t)= \int_{-\infty}^{+\infty} h(t-\hat{t})\,x_\textit{in}(\hat{t})\cdot d\hat{t}
    \label{eq:conv_impuls}
\end{equation}
The practical relevance of the impulse response stems from the fact that it can be measured operationally, in the limit of exciting the system with an approximate impulsive source.

Factoring the numerator and denominator polynomials of $H(s)$ into their roots yields the pole‑zero‑gain (PZ‑G) representation, which exposes the fundamental structure of the system dynamics. 
In this form, the complex-plane locations where the transfer function diverges (its poles) and where it vanishes (its zeros) are explicitly identified:

\begin{equation}
    H(s)  = K\cdot\frac{\prod_{j=0}^{m} (s-s_\textit{zj}) }{\prod_{k=0}^{n} (s-s_\textit{pk}) }
    \label{eq:P-Z-G}
\end{equation}

where $s_{zj}$ are the (complex‑valued) zeros, $s_{pk}$ are the poles, and $K$ is a real gain factor given by the ratio of the leading coefficients in Eq.~\ref{eq:TF_Laplace}. The pole-zero constellation is itself a complete description of the system's dynamics as it faithfully preserves the algebraic structure of $H(s)$. 

To guarantee physical realizability the high-frequency gain must be finite, a condition known as properness. This imposes that the number of zeros cannot be greater than the number of poles ($m \leq n$).

This representation is particularly advantageous when analyzing cascaded systems. For a series connection of $l$ LTI subsystems, the overall transfer function is simply the product of the individual transfer functions:

\begin{equation}
    H_\textit{tot}(s) =H_1(s)\cdot H_2(s)\cdots H_l(s) = \prod_{k=1}^{l} {H_k(s)} 
\label{eq:cascade}
\end{equation}

The poles and zeros of the combined system are the unions, with multiplicities, of the poles and zeros of each subsystem. The PZ-G representation therefore enables modular analysis and design, as the overall pole-zero constellation is obtained directly by aggregating those of the constituent stages.

When dealing with parallel modular combinations, which involve addition and subtraction instead of multiplication, a partial-fraction decomposition is more suitable. For a system with distinct poles, this yields an additive expansion:

\begin{equation}
    H_{tot}(s)  = \sum_{k=0}^{m} \frac{r_k}{(s-s_\textit{pk}) }
    \label{eq:residues}
\end{equation}
where each pole $s_{pk}$ contributes a first-order term wighted by the complex residue $r_k$. 
Because the Laplace transform and its inverse are linear but do not preserve multiplication, this additive form is particularly convenient for computing the combined impulse response:

\begin{equation}
    h_{tot}(t) = u(t) \cdot\sum_{k=0}^{m} r_k e^{s_{pk} t}
    \label{eq:impulse_sum}
\end{equation}
where $u(t)$ is the unit step function enforcing causality: in the convolution integral Eq.~\ref{eq:conv_impuls}, the output at any instant depends only on current and past values. 
Each exponential term is an eigenfunction of the underlying differential equation, Eq.~\ref{eq:ct-lti_diff}, and the residues $r_k$ determine the amplitude and phase contribution of each mode. 
Exponential and oscillatory functions of the form $e^{st}$ are invariant under differentiation up to a scaling factor; this eigenfunction property allows systematic construction of complete solutions to linear differential equations through superposition. This eigenfunction-perspective is central to the analysis developed in the following sections.

For real systems, complex poles and their residues occur in conjugate pairs, ensuring that the impulse response remains real-valued.
Eq.~\ref{eq:impulse_sum} formalizes how zeros, which get implicitly encoded into the residues, and the poles shape the time-domain response. The poles' location in the S-plane (root locus) determines whether its associated mode exhibits oscillation (non-zero imaginary part), exponential decay (negative real part), or exponential growth (positive real part). 
Exponential growth is excluded for stable physical systems, but it does not describe unphysical behavior. In unstable configurations, such as a positive feedback loop unchecked by stronger negative feedback, the system's response grows until physical saturation occurs and the transfer function model breaks down as the system no longer complies with LTI assumptions.   

The modal decomposition of Eq.~\ref{eq:impulse_sum} provides insight into the system's behavior and  will serve as the template for the discrete-time case, where exponential modes are replaced by geometric sequences. Isolating the contribution of repeated and complex-conjugate poles highlights specific modal behaviors: time-polynomial factors for the former and oscillatory terms for the latter. For $q$ distinct pole frequencies, each with multiplicity $n_q$, the associated contribution to the impulse response becomes:

\begin{equation}
    h_\textit{repeat}(t) = \sum_{j=1 }^q e^{s_{pj} t}\cdot u(t) \cdot\left( \sum_{k=1}^{n_q} c_{jk} \frac{t^{k-1}}{(k-1)!} \right) 
\label{eq:h_repeat}    
\end{equation}

where the coefficients $c_{jk}$ generalize the role of residues to the case of multiple poles.

Complex poles almost exclusively appear in conjugate pairs and can therefore be combined into real-valued sinusoidal terms:

\begin{equation}
    h_\textit{conj}(t) = \sum_{\text{j=1}}^{n} 2\rho_j e^{\sigma_j t} \cos(\omega_j t + \phi_j) \cdot u(t).
\end{equation}
where $\sigma_j$ and $\omega_j$ are the real and imaginary parts of the $j$‑th conjugate pole pair, and $\rho_j$ and $\phi_j$ are the magnitude and phase of the corresponding residues. For purely imaginary poles, $\sigma_j = 0$), the otherwise damped response reduces to sustained oscillations at angular frequency $\omega_j$.

Evaluating the Laplace variable along the imaginary axis, $s = i\omega$, restricts the eigenfunctions $e^{st}$ to purely oscillatory modes; this implies that an LTI system can only produce a scaled and phase-shifted version of a sinusoidal input.
This specialization leads directly to the frequency response interpretation of the transfer function:

\begin{equation}
    H(i\omega) = \frac{\sum_{j=0}^{m} b_j \cdot (i\omega)^j }{\sum_{k=0}^{n} a_k \cdot(i\omega)^k }
    \label{eq:freq_response}
\end{equation}

From Eq.~\eqref{eq:impulse_sum}, each pole at $s_{pk}$ contributes a modal frequency $\omega_k$ and a damping factor $\sigma_k$; hence the steady-state gain ad a given frequency $\omega$ depends on the distance from the imaginary components of the poles.

In terms of practical limitations, physical systems cannot provide infinite outputs and therefore gains. This precludes the faithful realization of ideal integrators ($H(s) \propto s^{-1}$) and differentiators ($H(s) \propto s$) in practice.
Another inherent limitation concerns achievable bandwidth: even a finite gain cannot be sustained at arbitrary frequencies, as unavoidable parasitic elements  (such as capacitance and inductance for analog electronic systems) always introduce a cutoff frequency. These dynamics restrict the gain at high frequencies and effectively bound the maximum operable frequency. This limitation also distorts signals with sharp transitions, whose high-frequency components are attenuated. The dominant frequency associated with a transition time $t_{\text{trans}}$ (measured between $10\%$ and $90\%$ amplitude levels) can be approximated as : 

\begin{equation}
    \omega_{\text{max}} \approx 2\pi \cdot \frac{0.35}{t_{\text{trans}}}.
\end{equation}

While bandwidth constraints inevitably limit the faithful reproduction of fast signal components, they also serve a critical purpose: attenuating high-frequency noise, most notably the ubiquitous thermal noise arising from random charge-carrier agitation. Thus, the same dynamics that bound a system's response simultaneously endow it with intrinsic noise-rejection capabilities.

\section{First-Order Continuous-Time Dynamics}

irst‑order transfer functions are characterized by numerator and denominator polynomials of degree at most one. In circuit theory, these transfer functions emerge directly from the constitutive relations of capacitors and inductors, which describe the derivative relationship between their state variables:

\begin{equation}
    \begin{cases}
        i_\textit{C}(t) = C \cdot \dfrac{dv_\textit{C}(t)}{dt} \\[8pt]
        v_\textit{L}(t) = L \cdot \dfrac{di_\textit{L}(t)}{dt}
    \end{cases}
\end{equation}

Upon Laplace transformation, these relations yield the corresponding $s$-domain impedances, which embody integration and differentiation:

\begin{equation}
    \begin{cases}
        H_\textit{C}(s) = \dfrac{V_\textit{C}(s)}{I_\textit{C}(s)}=Z_\textit{C}(s)=\dfrac{1}{s \cdot C} \\[8pt]
        H_\textit{L}(s) = \dfrac{V_\textit{L}(s)}{I_\textit{L}(s)}=Z_\textit{L}(s)=s \cdot L
    \end{cases}
\end{equation}
With voltage and current assigned as output and input, respectively, these component impedances take the form of ideal integrator (pole at $s=0$) and ideal differentiator (zero at $s=0$) transfer functions. In practice, however, physical components are subject to saturation arising from the finite electromagnetic fields they can sustain, which ultimately limits the achievable current and voltage. This restricted operating region gives way to nonlinear behavior beyond which the LTI model no longer applies.

\begin{figure}
    \centering
    \includegraphics[width=0.5\textwidth]{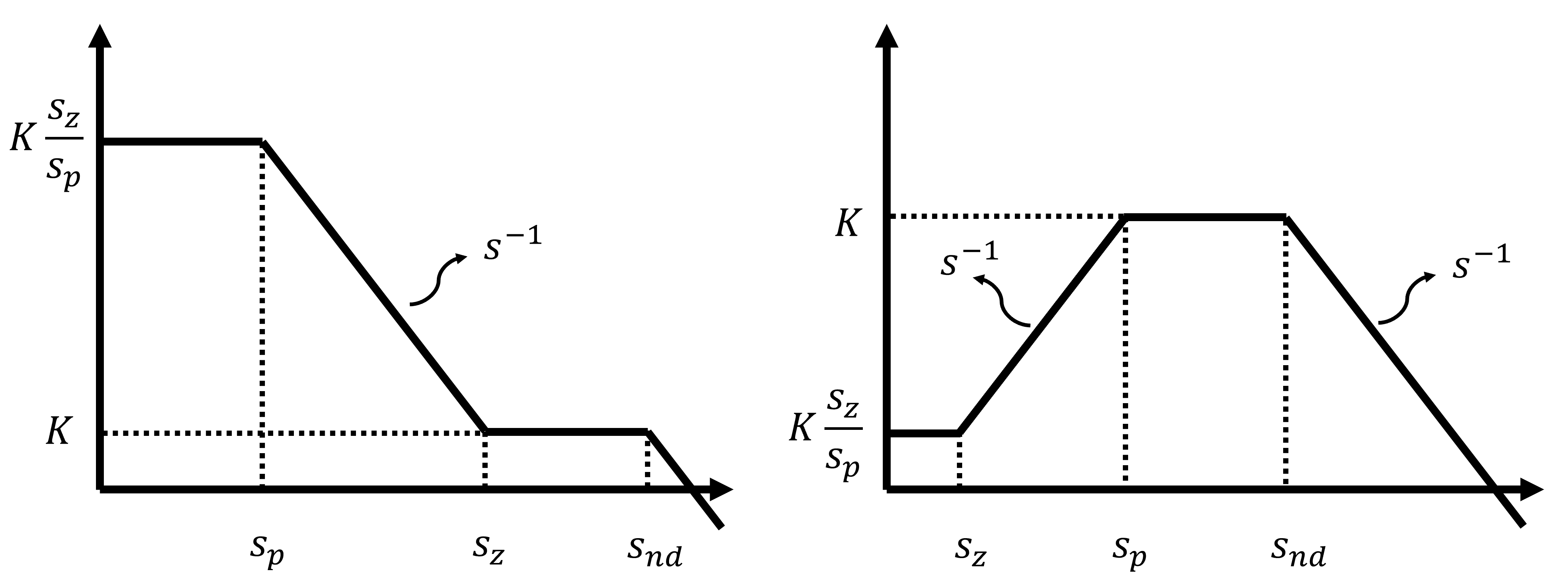}
    \caption{Approximated Lapace-domain transfer functions illutrating filter characteristics influenced by parasitic physical elements. Here $s_p$ denotes the first-order pole, $s_z$ the first order zero, and $s_{nd}$ the additional non-dominant pole that elevates the system to second-order behavior.}
    \label{fig:H_s_pole_zero}
\end{figure}

Additionally physical components exhibit distributed parasitic elements, like parallel capacitance and series inductance, which introduce additional poles and zeros respectively. Under a first-order approximation, these effects result in the transfer function:
\begin{equation}
    H_{phys}(s) = K\cdot\frac{s-s_z}{s-s_p} \cdot \frac{1}{s-s_{nd}}
    \label{eq:diff_tf}
\end{equation}
Where $s_{nd}$ is the non-dominant pole responsible for properness and 
which approximates an integrator for $s_p \ll s_z $ and a derivator when $s_z \ll s_p $, but only at for frequencies $s\in[s_z,s_p]$ and sufficiently far from the endpoints as depicted in Fig.~\ref{fig:H_s_pole_zero}.

Next, low-pass and high-pass filters are recalled to provide a reference for their DT counterparts, which are discussed in later sections.

\subsection{Low Pass Filter}
To keep input and output signals within the same electrical domain, a partitioning network, such as a voltage divider, can be implemented by introducing an additional impedance. Including a linear element like a resistor preserves the order of the resulting transfer function while enabling straightforward signal conditioning.

For example, a first‑order low‑pass filter in voltage mode is realized by the series combination of a resistor $R$ and a capacitor $C$. Combining their constitutive relations yields the transfer function:

\begin{equation}
    H_{\text{LPF}}(s) = \frac{1}{1+sRC} = \frac{1}{\tau} \frac{1}{s + \frac{1}{\tau}}
    \label{eq:low_pass_filter}
\end{equation}

where $\tau=RC$ is the circuit time constant which corresponds to the real pole $s_p=-1/\tau$. At frequencies well below the cutoff the gain approaches unity reflecting the diminishing current through the capacitor. 

The integrating action of the low-pass filter is revealed through the convolution integral with its impulse response: 

\begin{equation}
    \begin{cases}
        h_\textit{LPF}(t) = \mathcal{L}^{-1} \left[ \frac{1}{\tau}\frac{1}{s+\frac{1}{\tau}}\right]=\frac{1}{\tau}e^{-\frac{t}{\tau}}\cdot u(t) \\[8pt]
        x_{\textit{out}}(t) = \frac{1}{\tau} \int_{-\infty}^{t} x_{\textit{in}}( \hat{t}) \cdot  e^{-\frac{t-\hat{t}}{\tau}} \, d\hat{t} 
    \end{cases}
    \label{eq:lpf_output}
\end{equation}

where the unit step enforces causality by restricting the upper limit of integration. The filter output is thus the causal integral of the input weighted by an exponential decay that determines the domination of recent contributions over older ones. Beyond $4.6~\tau$, the residual contribution falls below $1\%$ of the original value. Effectively, the low‑pass filter performs a weighted moving average with an integration window on the order of $\tau$.
By contrast, an ideal integrator assigns equal weight to all past inputs:

\begin{equation}
    \begin{cases}
       h_\textit{integ}(t) = \mathcal{L}^{-1} \left[ \frac{1}{s} \cdot 1\right]=\displaystyle \int_{-\infty}^{t} \delta(\tilde{t})\cdot d\tilde{t}=u(t) \\[8pt]
       x_\textit{out}(t) = \displaystyle \int_{-\infty}^{+\infty} x_{in}(t-\hat{t})  u(t)\cdot d\hat{t} =\int_{-\infty}^{t} x_{in}(\tilde{t})  \cdot d\tilde{t}
    \end{cases}
\end{equation}
accumulating the entire history of the input without attenuation.

Although a pole formally causes the denominator to vanish in the $s$-domain, evaluating the frequency response along the imaginary axis does not produce divergence. This follows from the fact that physically realizable, stable systems admit only poles with negative real part; hence, as $s$ is restricted to the imaginary axis, it never coincides with a pole. In the frequency response $H(\omega)$, a pole instead manifests as a finite cutoff frequency, where the amplitude is attenuated by $3~dB$ relative to its passband value rolls off smoothly and linearly beyond this point: 
\begin{equation}
    \left| H_{\textit{LPF}}(s=i\omega) \right|_{\omega= \frac{1}{\tau}}= \left|\frac{1}{\tau}\frac{1}{\sqrt{\omega^2 + \frac{1}{\tau^2}}} \right|_{\omega= \frac{1}{\tau}}=\frac{1}{\sqrt{2}}
    \label{eq:LPF_freq_w}
\end{equation}

Another relevant non-ideality is the finite propagation delay $\tau_d$ inherent to physical systems and arising from finite signal propagation velocity, even under the lumped-component approximation. This delay enforces strict causality by further restricting the upper integration limit to $t-\tau_d$ and introduces a phase-lag factor in the transfer function:

\begin{equation}
    H_{\text{LPF,phys}}(s) = \frac{e^{-s\tau_d}}{1 + s\tau}.
\end{equation}

Notably, the exponential factor is not a rational function; consequently, the delay term introduces neither finite poles nor zeros into the transfer function. Its effect is confined to a frequency-dependent phase shift and the enforcement of strict causality, while the essential filtering functionality remains otherwise unaltered.

This correction is typically negligible and often omitted in first‑order lumped models, particularly at frequencies where $\omega\tau_d \ll 1$. However, the fact that the output cannot respond instantaneously to input changes becomes significant in high‑speed applications or distributed implementations.

\subsection{High Pass Filter}

Interchanging the positions of the resistor and capacitor in the passive RC network yields a first‑order high‑pass transfer function:
\begin{equation}
    H_{\textit{HPF}}(s)=  \frac{s}{s + \frac{1}{\tau}} = 1 -\frac{1}{\tau}\frac{1}{s+\frac{1}{\tau}}
    \label{eq:high_pass_filter}
\end{equation}
with the $\tau=RC$ inducing the same pole as in Eq.~\ref{eq:low_pass_filter}. This transfer function corresponds to the generic first‑order form of Eq.~\ref{eq:diff_tf} with $K=1$, a finite zero at $s_z=0$ a single real negative pole. At frequencies well above the cutoff the gain approaches unity reflecting the fact that the capacitor's impedance becomes negligible. The decomposition above reveals that a high-pass response is equivalent to subtracting the low-pass from the identity transfer function. 
Unlike for the low-pass filter, whose integrating action follows directly from the integration of convolution, the differentiating action of the high-pass characteristic is encoded more subtly:

\begin{equation}
    \begin{cases}
        
    h_\textit{HPF}(t) = \mathcal{L}^{-1} \left[ 1-\frac{1}{\tau}\frac{1}{s+\frac{1}{\tau}}\right]=\delta(t)-\frac{1}{\tau}e^{-\frac{t}{\tau}}\cdot u(t) \\[8pt]
    x_{\textit{out}}(t) =
    x_{in}(t)-\frac{1}{\tau}\int_{-\infty}^{t} e^{-\frac{t-\hat{t}}{\tau}} \cdot x_{in}(\hat{t})\, d\hat{t}
    \end{cases}
    \label{eq:impuls_resp_hpf}
\end{equation}
As suggested by the decomposition in Eq.\ref{eq:high_pass_filter}, the Dirac's delta
extracts the instantaneous value of the input, while the exponential term integrates past inputs with a decaying weight. Subtracting the moving average from the instantaneous signal, preserves rapid variations while attenuating the slower baseline components. 
By contrast the ideal differentiator is characterized by:

\begin{equation}
    \begin{cases}
       h_\textit{diff}(t) = \mathcal{L}^{-1} \left[ s \cdot 1\right]=\delta'(t) \\[8pt]
       x_\textit{out}(t) = \displaystyle \int_{-\infty}^{+\infty} \delta'(t-\hat{t})\,x_{in}(\hat{t})\cdot d\hat{t} = x'_{in}(t).
    \end{cases}
    \label{eq:diff_impulse_resp}
\end{equation}
where the doublet distribution, $\delta'(t)$, under the action of integration, extracts the derivative of the input. 

Strictly speaking, the conventional definition of the derivative is non-causal as it depends on future values, even if only under a limit. For continuous signals, causality can be preserved by adopting a backward-time convention, since the left and right limits then coincide.

The approximation of the doublet with $h_{HPF}(t)$ relies on the fact that the eigenfunctions of linear differential operators are, by definition, proportional to their derivatives. The scaling factor, representing the rate of change, is estimated by subtracting the moving average. This yields an effective approximation of the derivative, provided the high‑pass cutoff frequency well above the signal's spectral content. Otherwise, the moving average lags behind and fails to faithfully track the instantaneous evolution. 
The cutoff frequency introduced by the pole in Eq.~\eqref{eq:high_pass_filter} cannot be eliminated; it is, in fact, required for stability, as it bounds the maximum achievable gain and thereby ensures physical realizability. 

As in the low‑pass case, at the cutoff point the frequency response exhibits a $3~dB$ attenuation relative to the pass-band value. Below cutoff the response transitions from a constant-gain pass-band to a region where the gain becomes linearly proportional to frequency (derivator region):

\begin{equation}
   \left| H_{\textit{HPF}}(s=i\omega) \right|_{\omega= \frac{1}{\tau}}= \left|\frac{\omega}{\sqrt{\omega^2 + \frac{1}{\tau^2}}} \right|_{\omega= \frac{1}{\tau}}=\frac{1}{\sqrt{2}}
    \label{eq:hpf_ampl}
\end{equation}

In addition to parasitic poles that roll off the pass-band gain at higher frequencies, physical systems also exhibit a finite propagation delay between input and output:

\begin{equation}
    H_{\text{HPF,phys}}(s) = \frac{1}{\tau}\frac{s}{1 + s\tau}\cdot\frac{1}{\tau_{nd}}\frac{1}{1 + s\tau_{nd}} \cdot e^{-s\tau_d}
\end{equation}

Finally, a band-pass filter and  its reciprocal, the band-stop, can be implemented by
cascading of high-pass and a low-pass sections, provide their poles and zeros are placed sufficiently far apart to prevent meaningful interaction. The resulting transfer function will then be the product of terms like Eq.~\ref{eq:high_pass_filter} and Eq.~\ref{eq:low_pass_filter}, yielding at least an order two.

\section{From Continuous-Time Signals to Discrete-Time Sequences}
This section establishes how sampling links CT signals $x(t)$ to DT sequences $x[n]$ and shows how this relationship leads to the discrete-time Fourier transform (DTFT) and, under periodic sampling, to the discrete Fourier transform (DFT).

DT LTI systems implement signal-conditioning functions through the basic arithmetic operations of addition, subtraction, and scaling—applied to samples of a time sequence. To perform DT integration and differentiation, as well as their combination into specific filtering responses, access to past samples is required. This is achieved using delay elements which effectively enables memory: building arbitrary chains of delays grant future computations access to past samples.
In digital implementations, this delay is typically realized with clocked registers, each storing a sample for one clock period. Analog sampled-data systems, by contrast, often employ capacitors to store analog voltages dynamically, using multi-phase switching to separate sampling and processing phases. 
Regardless of the implementation choice, the same mathematical models are applicable. The simpler one is the generic feed‑forward structure  where the current output only depends on past inputs:

\begin{equation}
    x_\textit{out}[n] = \sum_{j=1}^m{b_j\cdot x_\textit{in}[n-j]}
    \label{eq:fir_sequence}
\end{equation}

A richer class of systems is obtained by incorporating feedback, a signal path from the output back to the input, which introduces dependence on past output samples and results in the general recursive sequence:

\begin{equation}
    \sum_{k=0}^l{a_k\cdot x_\textit{out}[n-k]}  = \sum_{j=1}^m{b_j\cdot x_\textit{in}[n-j]}
    \label{eq:iir_sequence}
\end{equation}

Unlike the CT case, where an ideal impulse (Dirac delta) is by definition unrealizable,requiring infinite bandwidth and amplitude, its DT counterpart is straightforward to implement, consisting merely of a single non-zero sample in a sequence.
Feed-forward systems, described generally by Eq.~\ref{eq:fir_sequence}, produce at most $m$ non-zero output samples in response to an impulse and are therefore classified as finite-impulse-response (FIR) filters. 
The recursive form given in Eq.~\ref{eq:iir_sequence} yields an impulse response that, in principle, extends indefinitely because of feedback; systems of this type are therefore classified as infinite‑impulse‑response (IIR) filters. Stability requirements, however, constrain the response to decay exponentially over time. While a decaying exponential approaches zero only asymptotically, practical implementations exhibit finite responses as the signal eventually falls below the resolution limit of the system (whether determined by the finite precision of a digital representation or by the noise floor in an analog one).

\subsection{The Z-transform}

The transfer function of a DT system is modeled using the Z-transform, a mathematical tool analogus to the Laplace transform for CT systems. The Z-transform weights each element of a time sequence by a complex term $z^{-n}$. For finite duration sequences, it is convenient to adopt the unilateral formulation and choosing the time index so that the first non-zero sample occurs at $n=0$:

\begin{equation}
    X_\textit{in}(z) = \mathcal{Z}\{x_\textit{in}[n]\} = \sum_{k=0}^{\infty} x_\textit{in}[k] \cdot z^{-k}
    \label{eq:Z_transform}
\end{equation}

If the sequence is of finite length, there exists an index $m$ such that $x_\textit{in}(n)=0$ for all $n \geq m$. In this case, the sum in Eq.~\ref{eq:Z_transform} contains only a finite number of non-zero terms. The implication is that the function $X_{in}(z)$ is well-defined across the entire complex plane. 
Essentially, the Z-transform encodes the information about the input signal into a  mapping of the complex plane $X_\textit{in}(z):\mathcal{C}\rightarrow\mathcal{C}$. Alternatively, the map can be decomposed into polar components, two real-valued functions which provide the amplitude and phase of the complex image.  
Expressing $z$ in polar form makes explicit the role of the magnitude $\rho$, which geometrically scales successive samples, and the role of the phase $\theta$, which tracks the temporal ordering:

\begin{equation}
\begin{split}
    X_\textit{in}(z) =  x_\textit{in}[1] \cdot \rho^{-1}e^{-j\theta}+x_\textit{in}[2] \cdot \rho^{-2}e^{-j2\theta}+\cdots \\ 
    \label{eq:Z_transform2}
\end{split}
\end{equation}

The relationship between complex phase and time‑delay follows directly from the structure of periodic signals. For a sinusoid of angular frequency $\omega$, the instantaneous phase evolves linearly in time: $\theta(t)=\omega_0 t+\theta_0$. Using Euler’s formula, the cosine function can be written as the real part of a complex exponential:

\begin{equation}
    \mathit{cos}\left(\omega_0 (t \pm \Delta t)+\theta_0\right) = \operatorname{Re} \left[ e^{i(\omega_0 t+\theta_0)}\cdot e^{\pm i\omega_0 \Delta t}\right]
\end{equation}

This relation, highlights how a time delay becomes equivalent to a phase shift which can be obtained by multiplication of a pure phase in the complex exponential representation.

The connection between complex phase and time‑delay extends naturally from periodic signals to arbitrary discrete‑time sequences, which can be constructed by sampling a time-continuous signal at a fixed rate $F_s$:

\begin{equation}
    x_\textit{in}[n]=x_\textit{in}(t)\big|_{t=\frac{n}{F_s}}
    \label{eq:time_seq}
\end{equation}
Here the delay between consecutive samples is by construction equal to one sampling period $T_s=1/F_s$.
In the Z-transform of the sequence the delay information is retained by the factor $z^{-k}$ which accounts for the cumulative phase lag corresponding to a delay of $k$ samples. For a sinusoidal input with angular frequency $\omega_0$:

\begin{equation}
    \angle (z^{-k})=e^{i\Delta \theta_k} =e^{-i\omega_0 \Delta t_k} =  e^{-i\frac{\omega_0}{F_s} k}
    \label{eq:phase_shift}
\end{equation}

For a generic input, $\omega_0$ denotes the angular frequency of the signal itself, if periodic, or of its harmonic components in the non-periodic case.

By contrast, the modulus of the complex variable $\rho^{-k}$ governs the convergence of the series in Eq.~\eqref{eq:Z_transform}. For the Z‑transform to be well‑defined, the infinite series must converge to a finite, non‑trivial value. Convergence depends critically on the nature of the signal. For periodic or infinite‑duration signals, the region of convergence (ROC) always excludes the interior of the unit circle ($|z| < 1$), because their geometric factors would grow boundless with $k$.

This condition mirrors the stability requirement for $s$‑domain poles, which mandates negative real parts. Indeed, the correspondence between exponentials and geometric series implies that $\operatorname{Re}(s_p) \leq 0$ translates to $|z_p| \leq 1$ in the discrete domain.

For finite‑length sequences, the infinite series reduces to a polynomial in $z^{-1}$, which converges for all finite $z$ except the origin. Points at infinity are likewise excluded from the ROC, as they would correspond to a zero that trivializes the transform.

\subsection{Discrete-Time Fourier Transform} 
A frequency-domain representation that provides the spectral information of a signal is obtained from the Z‑transform by restricting the complex variable to the unit circle:

\begin{equation}
    X_{\text{in}}(z)\big|{z=e^{i\theta}} = X_{\text{in}}(\theta) = \sum_{k=0}^{\infty} x_{\text{in}}[k] \cdot e^{-i\theta k}
    \label{eq:DTFT}
\end{equation}

where the complex variable now traverses a periodic sequence of samples along the unit circle. Geometrically, each point on the unit circle encodes the phase accumulation associated with a given time delay. Tracking how this phase evolves with time is equivalent to performing a frequency analysis.
In fact, interpreting the index $k$ as discrete time measured in samples, rather than as a dimensionless integer, establishes the relationship between the complex phase $\theta$ and the discrete angular frequency $\omega_d$, expressed in radians per sample:

\begin{equation}
    \theta =\omega_d \cdot 1= \frac{\omega}{F_s}
    \label{eq:discr_freq}
\end{equation}

where $F_s$ is the sampling frequency. Equation~\eqref{eq:discr_freq} formalizes the correspondence between a continuous‑time frequency of interest $\omega$ and the resulting uniform angular spacing of samples on the unit circle.

The ratio $2\pi / \omega_d$ then gives the number of samples taken per period of the original continuous‑time signal $x_{\text{in}}(t)$—or, equivalently, per period of its harmonic component at frequency $\omega$.

To uniquely determine a sinusoid, up to an integer multiple the sampling frequency, at least three independent points are needed. This allows the amplitude, frequency and initial phase to be resolved. Having only two equally spaced samples per period produces a repeating sequence that admits an infinite family of amplitude-phase pairs. 
With slightly more than two samples per period ($\omega<\pi F_s$) the sinusoid is sampled at distinct phase points in each period, enabling unique identification.

A discrete sinusoid identifies a discrete spectral component of the transformed signal, whose amplitude and phase values are obtained by evaluating the DTFT of Eq.~\ref{eq:DTFT} at the discrete frequency given by Eq.~\ref{eq:discr_freq}.

The DTFT in Eq.~\ref{eq:DTFT} is a continuous and periodic function of the discrete angular frequency. The complex exponential introduces a periodicity of $2\pi$ radians, causing the frequency axis to repeat indefinitely. In addition, the symmetries properties of trigonometric functions introduce additional structure. In fact, 

\begin{equation}
   e^{i(\theta+\pi\cdot n)}= (-1)^{n} \cdot e^{i\theta} 
\end{equation}
implies that shifting the frequency by integer multiples of $\pi$ either preserves or flips the sign of the complex exponential depending on the parity of $n$. As a result, the magnitude spectrum repeats identically over intervals $[\pi \cdot n_{\text{even}}, \pi \cdot (n_{\text{even}}+1)]$ and appears mirrored over intervals $[\pi \cdot n_{\text{odd}}, \pi \cdot (n_{\text{odd}}+1)]$.
This effect is called aliasing and restricts the meaningful analysis of discrete‑time signals to the Nyquist range, $\omega_d \in[0,\pi]$, corresponding to continuous frequencies $f\in[0,F_s/2]$. Beyond this range, it is no longer possible to establish a unique mapping between a continuous‑time signal and its sampled sequence: multiple higher‑frequency signals become indistinguishable from lower‑frequency ones when sampled at a given rate.
For instance, a discrete sinusoid sampled at $F_s=\omega/2\pi$ is identical to a constant sequence, as it is sampled once per period always at the same phase point.



The DTFT sum in Eq.~\ref{eq:DTFT} converges absolutely only for finite-length or exponentially decaying sequences. For periodic signals like sinusoids and their linear combinations, the DTFT can still be defined, albeit in  a generalized form where convergence is achieved under the action of test functions. 
In fact for a discrete sinusoid the DTFT becomes:

\begin{equation}
   \sum_{k=0}^{\infty} cos(\omega_0 k) \cdot e^{-i\omega_{d} k}=\pi\sum_{k=-\infty}^{\infty} \left[\delta(\omega_d\pm\omega_0-2\pi k) \right]
    \label{eq:DTFT_cos}
\end{equation}

which constitutes a collection of spectral lines at the fundamental frequency $\omega_0$ and its harmonics.

\subsection{Discrete Fourier Transform}

Because the DTFT is defined for an infinite-length sequence, practical implementations must truncate the signal to a finite length $N$ and instead compute the discrete Fourier transform (DFT), which yields a discrete sequence of the same length:

\begin{equation}
     X_\textit{in}(\omega_k) = \tilde{X}_\textit{in}(e^{-in\omega_k}) = \sum_{k=0}^{N-1} x_\textit{in}[k] \cdot e^{-in\omega_k}
    \label{eq:DFT}
\end{equation}
where $ \tilde{X}_{in}(z)$ is the finite approximation of the Z-transform and the discrete frequency points are given by:

\begin{equation}
\omega_k=2\pi kF_s/N \quad \text{for} \quad k = 0,1, \dots, N-1
    \label{eq:dft_freq}    
\end{equation}
This formulation highlights that $\omega_k$ corresponds to a uniform sampling of the continuous frequency variable \footnote{Although called the discrete frequency, $\omega_d$ is in fact a continuous variable: it is simply the continuous angular frequency $\omega$ normalized by the sampling rate $F_s$. The descriptor "discrete" properly applies to the time-domain sequence, not to the frequency variable itself.} $\omega_d$. The DFT thus represents a uniformly sampled version of the DTFT at $N$ points over the interval $[0, 2\pi F_s)$.

The Nyquist frequency corresponds to the index $k = N/2$, and the spectral resolution—determined by the spacing between consecutive frequency samples—is given by the bin width $\Delta\omega = 2\pi F_s / N$.

The validity of Eq.~\eqref{eq:DFT} follows from the fact that any uniformly sampled, perfectly periodic sequence of length $N$ admits an exact decomposition into a finite set of sinusoids at the frequencies given by Eq.~\eqref{eq:dft_freq}. 
Equivalently, because discrete complex exponentials form an orthonormal basis for all periodic sequences, the set of frequencies that uniquely represent the signal is entirely determined by the sequence length $N$ and the sampling rate $F_s$.

The periodicity of the time-domain sequence follows from the mathematical duality between sampling and periodic repetition in the time and frequency domains. Specifically, sampling the DTFT at $N$ equally spaced points implicitly converts the finite-length sequence into a periodic one, satisfying $x_{\text{in}}[N \cdot n] = x_{\text{in}}[N \cdot (n+1)]$. Finally, because both factors in the summation of Eq.~\eqref{eq:DFT} are periodic modulo $N$, the resulting sampled spectrum can likewise be repeated across the discrete frequency axis.

Evaluating the DTF for a discrete cosine yields:

\begin{equation}
    \sum_{n=0}^{N-1} \cos\!\left(\frac{2\pi k_0}{N} n\right) e^{-i \frac{2\pi k}{N} n}= \frac{N}{2}\delta[k\pm k_0-N]
    \label{eq:DFT_cos}
\end{equation}
where the indices $k$ can be extended modulo $N$ and $\delta[-]$ is the Kronecker delta which represents the sampled version of a Dirac's impulse. 
When the ratio between the signal frequency and the sampling frequency is non‑integer, the finite sequence does not capture an integer number of periods. Consequently the DFT’s implicit periodic extension introduces an artificial discontinuity at the boundaries due to $x_\textit{in}(N/F_s) \neq x_\textit{in}(0)$. This sharp transient contains high‑frequency components which do not always end up in the same location across the replicated spectra, therefore their aliasing into the fundamental Nyquist band induces the discontinuity energy to distribute over all frequency bins, a phenomenon called spectral leakage.
Mathematically, sampling a signal whose frequency is not an integer submultiple of the sampling rate is equivalent to filter its time-domain representation with a rectangular window that does not span an integer number of periods. This time-domain multiplication corresponds to a convolution in the frequency domain with the spectrum of the window, in this case, a cardinal sine (sinc) function. This operation produces the same spectral leakage previously described.   
A practical mitigation approach is to use sequences that encompass more periods of the signal. Then, the relative energy of the discontinuity decreases, as its contribution is averaged over more periods, yielding a more faithful spectral representation. Alternatively, extending the duration of window narrows the main lobe of its spectrum, thus, reducing the spectral leakage under convolution.

\section{Discrete-Time LTI Systems and the Z-Domain}
The role played by complex exponentials as eigenfunctions and modal decompositions in CT LTI systems is assumed by geometric sequences in the discrete-time domain, as this section will detail.
While the transfer function of a continuous‑time system, Eq.~\ref{eq:TF_Laplace},  is derived via the Laplace transform, discrete‑time systems are naturally characterized using the Z‑transform, leading to a transfer function of the form polynomial ratio of the type: 

\begin{equation}
H(z) = \frac{X_{\text{out}}(z)}{X_{\text{in}}(z)} = \frac{\sum_{j=0}^{m} b_j z^{-j} }{\sum_{k=0}^{n} a_k z^{-k} }
\label{eq:Z_trans_TF}
\end{equation}

In the DT domain, the infinitesimal operations of differentiation and integration are replaced by finite differences and accumulation respectively, providing a simple algebraic relation between the samples in the sequence. In the Z‑domain as well, transfer functions offer clear insight into frequency‑domain behavior and impulse‑response characteristics, while simplifying the analysis of interactions between subsystems. These features are essential for enabling systematic, modular design of discrete‑time systems. 
As for the CT case, strict causality imposes a backward definition of the derivatives. Therefore the time-discrete derivatives of a sequence $x[k]=x(t)|_{t=k\cdot T_s}$ are defined as: 

\begin{equation}
\begin{cases}
   x^{(1)}[k]=\frac{x[k]-x[k-1]}{T_s} \\[8pt]
   x^{(2)}[k]=\frac{x^{(1)}[k]-x^{(1)}[k-1]}{T_s}=\frac{x[k]-2x[k-1]+x[k-2]}{T_s^2}  \\[8pt]
   x^{(3)}[k]=\frac{x[k]-3x[k-1]+3x[k-2]-x[k-3]}{T_s^3}  \\[8pt]
   \cdots
\end{cases}   
\end{equation}

where $T_s$ is the inverse of the sampling rate $F_s$. Higher order derivatives are built recursively as differences of differences and can be synthetically expressed with the general formula:

\begin{equation}
    x^{(n)}[k] =\frac{1}{T_s^{n}} \sum_{m=0}^{n}(-1)^{m}\binom{n}{m} x[k-m]
\end{equation}

The approximation error introduced by the sampling operation can be derived from the backward Taylor expansion:

\begin{equation}
    x\big((k-1)T_s\big) = x(kT_s)+\sum_{n=1}^\infty\frac{(-T_s)^n}{n!}\cdot\frac{d^nx(t)}{dt^n}\bigg|_{t=kT_s}
\end{equation}

Isolating the first derivative and reordering shows how the error scales linearly with the sampling time:

\begin{equation}
    \frac{d}{dt} x(k  T_s) =\frac{x[k]-x[k-1]}{T_s} +O(T_s)
\end{equation}

Similarly, the DT approximation of the integration is based on the integrand expansion:
\begin{equation}
    x(t_2) = x(t_1)+\sum x^{(n)}(t_1) \frac{(t_2-t_1)^n}{n!}
    \label{eq:discrete_expansion}
\end{equation}
using $t_1=(k-1) T_s$ and $t_2=k Ts$, constrains the analysis to a single time-bin and yields:  

\begin{equation}
    \int_{t_1}^{t_2} x(\hat{t})\,d\hat{t} = x[k]\cdot T_s+\sum _{n=1}^{\infty} x^{(n)}[k]\cdot T_s
    \label{eq:discr_integ}
\end{equation}

Based on the number of terms retained in the discrete expansion of Eq.~\ref{eq:discrete_expansion}, integration over multiple samples can be approximated with different orders of accuracy. Using zero‑order interpolation yields the rectangular rule, whose global error scales linearly with the sampling period:

\begin{equation}
    \int_{p\cdot T_s}^{q\cdot Ts} x(\hat{t})\,d\hat{t}= \sum _{k=p}^{q-1} x[k]\cdot T_s+O(T_s)
    \label{eq:zero_order_accum}
\end{equation}

Including the first‑order term corresponds to linear interpolation between samples, giving the trapezoidal rule with a quadratic global error:

\begin{equation}
    \int_{p\cdot T_s}^{q\cdot Ts} x(\hat{t})\,d\hat{t} =  \sum _{k=p+1}^{q} \frac{x[k-1]+x[k]}{2} T_s+O(T_s^2) 
    \label{eq:1st_order_accum}
\end{equation}

In both implementations the integration is essentially an accumulation of samples, the accuracy is improved by simply improving the estimation of the height value which makes the rectangle rule exact.


\subsection{First-Order Discrete-Time Dynamics}
\subsubsection{Discrete low-pass filter}

The DT counterpart of differential equations that describe the state variables of a dynamical system are finite difference equations involving consecutive samples. In principle the direct conversion from continuous to discrete time with the convention $T_s=1$ entails:
\begin{equation}
    \frac{dx}{dt}(t)=a\cdot x(t) ~~\Rightarrow ~~ x[n]-x[n-1] = a\cdot x[n]
\end{equation}

however a simple coefficient conversion allows for equivalent formulations of difference equations:

\begin{equation}
    x[n] - x[n-1] =a-1=b
\end{equation}
and the recurrence relation: 
\begin{equation}
    x[n] = \frac{x[n-1]}{1-a} = c \cdot x[n-1]
    \label{eq:seq_difference_eq}
\end{equation}

The continuous exponential function, $e^{st}$, is defined as the eigenfunction of linear differential equations, where at each point the derivative is proportional  to the function itself. Similarly, the discrete exponential is defined as the eigenfunction of linear difference equations, where each sample is a scaled version of the preceding one, yielding the geometric sequence:

\begin{equation}
    x[n] = x[0] \cdot\prod_{k=0}^{n-1} c_k = x[0] \cdot c^{n}. 
    \label{eq:DT_exp}
\end{equation}

The base of the exponential, determines the dynamic behavior of the system. With $|c|<1$, the system is stable and the function is a decaying exponential. With $|c|>1$, the exponential growth indicates instability. The case of $c=1$ corresponds to marginal stability as $x[n]$ is a constant step; whereas $c=-1$ produces an indefinite sign alternation at the Nyquist frequency $2F_s$. When $c$ is complex the sequence exhibits proper oscillatory behavior as the argument of the polar representation, $c=\rho_c e^{i\theta_c}$, directly represents the phase increment per sample:

\begin{equation}
    x[n] = x[0] \cdot \rho_c^{n}e^{in\theta_c} 
\end{equation}

this implies that the angular frequency of the oscillation is simply $\omega_{osc}=\theta_c/T_s$ and the number of samples covering one oscillation period is $N=2\pi/\theta_c$. Whereas, the module $\rho_c$ accounts for the exponential envelope of the oscillation. 
In fact applying the Z-transform reveals that the parameter $c$ corresponds to the pole of system modeled by Eq.\ref{eq:seq_difference_eq}. Invoking causality ($x[n]=0$ for $n<0$) and noting that the recurrence holds only for $n>0$, the Z-transform of the sequence must be taken from $n=1$, which leaves the initial condition $x[0]$ as a free term:

\begin{equation}
    \mathcal{Z}(x[n])=\sum_{n=1}^{\infty}x[n]\cdot z^{-n} = X(z)-x[0]
\end{equation}
For the right side of the recurrence: 
\begin{equation}
    \mathcal{Z}(c  x[n-1])=c  \sum_{n=1}^{\infty}x[n-1]z^{-n}z^{\pm1}=c z^{-1}\cdot X(z)
\end{equation}
Then treating the initial condition as a discrete impulse $\delta[n]=x[0]=x_0$ is equivalent to considering a discrete system with a transfer function between input and output given by: 

\begin{equation}
    H_{LPF}(z) = \frac{x_0}{1-cz^{-1}}
\end{equation}

which represents a first-order system with a single pole at $z=c$. The requirement for stability, and hence convergence of the impulse response, is met when the pole lies inside the unit circle ($|c|<1$). From Eq.~\ref{eq:DT_exp}, the response to a unit impulse, $x_0=1$, is a causal, decaying discrete exponential:

\begin{equation}
    h[n] = c^{n}\cdot u[n]
    \label{eq:impls_resp_pole}
\end{equation}

where $u[n]$ denotes the discrete unit step equal to $0$ for negative $n$ and $1$ otherwise. As in the continuous case, the response to an arbitrary input sequence is obtained by convolution with the impulse response:

\begin{equation}
    x_{out}[n]=\sum_{k=-\infty}^{\infty}h[k]x_{in}[n-k]=\sum_{k=-0}^{\infty}c^k\cdot x_{in}[n-k]
\end{equation}

This expression corresponds to a weighted accumulation of past inputs, with an exponential weight that attenuates older samples, thus the characteristic behavior of a leaky integrator. This reveals how the local form of a first‑order difference equation Eq.~\ref{eq:seq_difference_eq} globally implements a discrete approximation of the ideal integration of Eq.~\ref{eq:zero_order_accum}.

\subsubsection{Discrete high-pass filter}

An ideal differentiator cannot be captured by a first‑order homogeneous difference equation of the form in Eq.~\ref{eq:seq_difference_eq}; instead it requires an explicit input‑output relation. The impulse response of an ideal discrete‑time differentiator is the discrete doublet, which corresponds to a sampled version of the continuous‑time derivative kernel in Eq.~\ref{eq:diff_impulse_resp}:

\begin{equation}
    h[n]=\delta[n]-\delta[n-1]
    \label{eq:id_diff_TF}
\end{equation}

Convolving this doublet with an arbitrary input yields the output:

\begin{equation}
    x_{out}[n] = \sum_{k=-\infty}^{\infty} h[k] \cdot x_{in}[n-k] = x_{in}[n] - x_{in}[n-1]
\end{equation}

where each output sample is simply the backward difference of the input sequence. Thus, the ideal differentiator is a memoryless two‑tap FIR filter, fundamentally distinct from the recursive (IIR) structure of the first‑order low‑pass or leaky‑integrator filters. 
It's transfer function, the Z-transform of the doublet (Eq.~\ref{eq:id_diff_TF}), is simply:

\begin{equation}
    H(z) = 1 - z^{-1}.
\end{equation}

The zero at $z=1$ corresponds to DC, since the phase term (i.e. the angular frequency) is always zero, and is responsible for attenuating low-frequency components. The recursive (IIR) implementation requires introducing a pole inside the unit circle to ensure stability and convergence. This produces a high-pass filter characteristic that approximates differentiation below the pole's cut-off frequency. Introducing a stabilizing pole corresponds to adding a term in the recurrence which is proportional to the delayed output as in Eq.~\ref{eq:seq_difference_eq}, yielding :

\begin{equation}
    x_{out}[n]=\left(x_{in}[n]-x_{in}[n-1]\right)+ax_{out}[n-1]
    \label{eq:HPF_recursion}
\end{equation}

For a unit-impulse input, the impulse response becomes a decaying alternating sequence: 

\begin{equation}
  h[n]= (a^{n}-a^{n-1})\cdot u[n]
  \label{eq:discr_hpf_imp_resp}
\end{equation}

where the unit step preserves causality. The impulse is transformed into a doublet, via delay and sign inversion, and then shaped by the exponential decay imposed by the recurrence. The difference in exponents explicitly captures the effect of the delay. 
The closed form of the response to an arbitrary input output is obtained by convolution with Eq.~\ref{eq:discr_hpf_imp_resp} yielding:

\begin{equation}
    x_{out}[n]=\sum_{k=-\infty}^{\infty}(a^{k}-a^{k-1}) u[k]\cdot x_{in}[n-k]
\end{equation}

Which becomes:

\begin{equation}
    x_{out}[n]=\sum_{k=0}^{\infty}a^{k} \cdot x_{in}[n-k]-\sum_{k=0}^{\infty}a^{k-1} \cdot x_{in}[n-k\pm1]
\end{equation}

Renaming the index $k-1=j$, shows how the output is the difference between the exponentially weighted averages of the input and its delayed version, resulting in a leaky derivator. 

\begin{equation}
    x_{out}[n]=\sum_{k=0}^{\infty}a^{k} \cdot x_{in}[n-k]-\sum_{j=0}^{\infty}a^{j} \cdot x_{in}[n-1-j]
\end{equation}

The stabilization of the IIR architecture is obtained by replacing the direct backward difference with a smoothed version, where the exponential weights are responsible for the averaging. 
The transfer function can be easily obtained by Z-transforming the recurrence of Eq-~\ref{eq:HPF_recursion}:

\begin{equation}
    H_{HPF}(z)=\frac{1-z^{-1}}{1-az^{-1}}
    \label{eq:hpf_iir_tf}
\end{equation}
where the zero at DC, $z=1$, is compensated by the low-pass filtering pole at $z=a$.

\subsection{Conversion from Laplace to Z-transform}

The systematic design of DT filters typically proceeds by translating desired frequency‑domain specifications, originally formulated in terms of natural (continuous‑time) frequencies, into the complex plane. This conversion is achieved by establishing a mapping between the Laplace, $s$, and the Z‑transform, $z$, variables. The fundamental link is given by the relationship

\begin{equation}
z^{-1} = e^{-s \cdot T_s} = \sum_{n=0}^{+\infty} \frac{(-s \cdot T_s)^n}{n!}
\end{equation}

which reflects the fact that a unit delay essentially represents a phase shift.
For practical implementation, the infinite series is often truncated to a rational approximation such as: 

\begin{equation}
    s \approx \frac{1}{T_s} \cdot (1 - z^{-1})
    \label{eq:s_approx_z}
\end{equation}

This first-order approximation, known as the backward‑Euler discretization rule, preserves stability by mapping the left‑half s-plane maps into the interior of the unit circle. It also provides a straightforward algebraic substitution 
for converting transfer functions. Alternative mappings such as the bilinear (Tustin) transform provide better frequency-warping behavior by using trapezoidal integration, at the cost of a more complex substitution.
A more practical and precise design approach treats the discrete‑time system as a modular cascade, factoring individual poles and zeros. Rearranging Eq.~\ref{eq:Z_trans_TF} yields the discrete‑time counterpart of the pole‑zero‑gain representation in Eq.~\ref{eq:P-Z-G}:

\begin{equation}
    H(z)  = K\cdot\frac{\prod_{j=0}^{m} (1-z_\textit{zj}z^{-1}) }{\prod_{k=0}^{n} (1-z_\textit{pk}z^{-1}) }
\end{equation}

The conversion between CT and DT poles and zeros follows the exact mapping:

\begin{equation}
    z_{p,z} = e^{s_{p,z} \cdot T_s} ~~\Leftrightarrow ~~  \sigma_{p,z} + i \omega_{p,z} = \frac{1}{T_S} \cdot \log(z_{p,z})
    \label{eq:pole_zero_map}
\end{equation}

This factorization preserves the structure of the original transfer function while enabling direct implementation as a cascade of arbitrary-order DT sections.

\subsection{Sampled-Data Filter Implementations}
The principal first-order filter architectures are presented in this subsection. Each system is synthesized from unit-delays and summing-subtracting elements. The order of a given filter is defined by the polynomial degree in its transfer function which includes all feedforward and feedback paths. To maintain strict causality and hardware compatibility, an external unit delay is sometimes introduced outside this path, thus without affecting the order of the filter.

\begin{figure*}
    \centering
    \includegraphics[width=\textwidth]{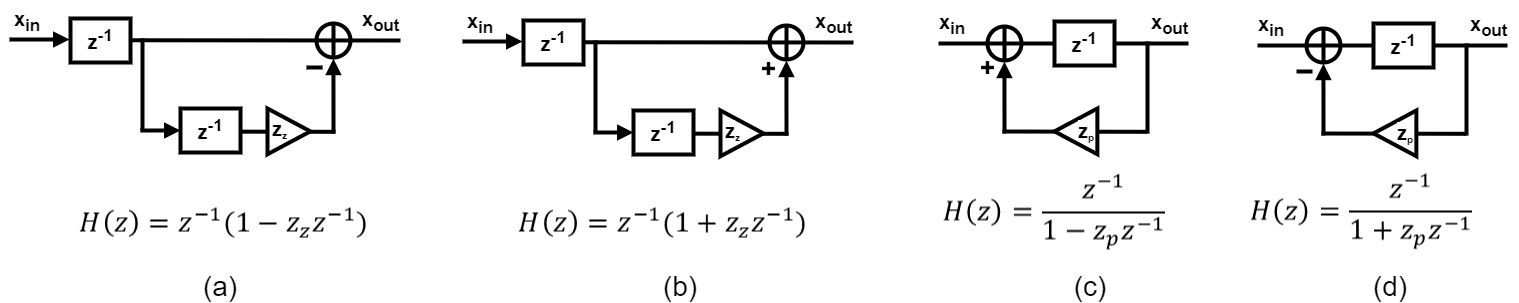}
    \caption{Block diagram of basic first-order DT LTI systems constructed from unit delay, amplitude scaling, and summation elements. (a) FIR differentiator with zero at $z_z$. (b) FIR two-sample moving average with a negative zero at $-z_z$. (c) IIR accumulator with pole at $z_p$. (d) IIR oscillator with a negative pole at $-z_p$.}
    \label{fig:block_diagram}
\end{figure*}

\begin{figure}
    \centering
    \includegraphics[width=0.5\textwidth]{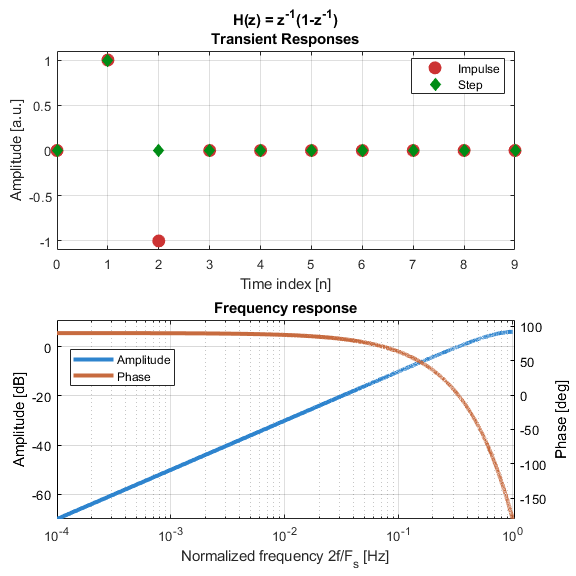}
    \caption{Transient (top) and frequency (bottom) response of the discrete differentiator. The impulse response is a discrete doublet which under convolution provides the difference between consecutive samples. The step response is a discrete impulse. The amplitude response shows a high-pass characteristic with a zero at DC $z=1$ and a maximum of $6~dB$ at Nyquist $z=-1$.}
    \label{fig:sys1_resp}
\end{figure}

The FIR architecture Fig.~\ref{fig:block_diagram}-(a) with a unitary amplitude coefficient $z_z=1$ constitutes a discrete differentiator (ideal discrete derivation). The corresponding transient responses in Fig.~\ref{fig:sys1_resp} shows, as expected, a doublet for the impulse response and an ideal impulse for the step response. Mapping the discrete normalized frequency to its continuous-time equivalent in hertz yields the effective frequency response of a system that samples a CT signal, filters it in the discrete domain, and reconstructs it to continuous time. The equivalent CT response
exhibits a maximum gain of $6~dB$ at the Nyquist frequency $f=F_s/2$, while the zero at DC ($z^{-1}=1$) provides the expected attenuation of the low-frequencies.

\begin{figure}
    \centering
    \includegraphics[width=0.5\textwidth]{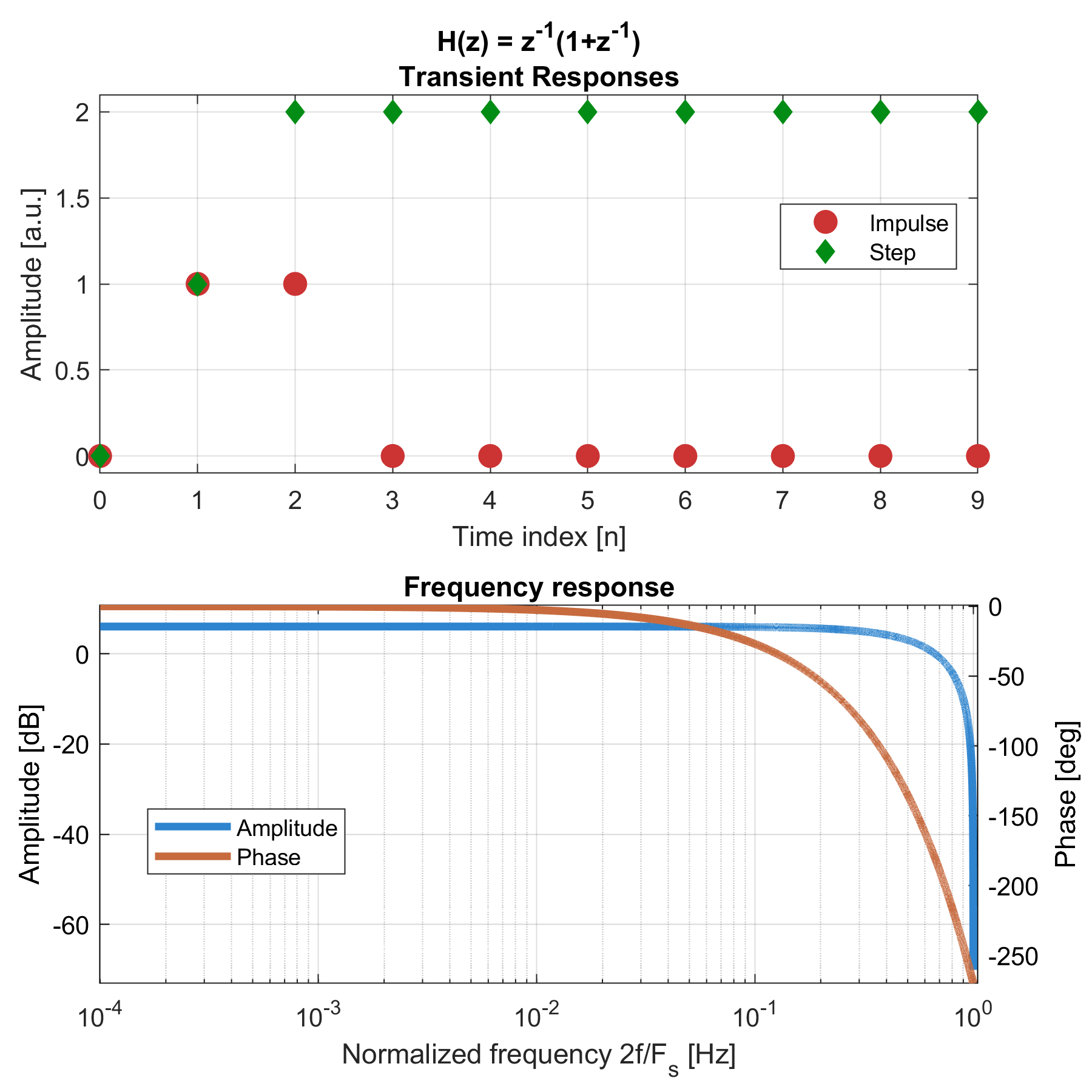}
    \caption{Transient (top) and frequency (bottom) response of the 2-sample moving sum. The impulse response is a 2-sample rectangular pulse  which under convolution provides the sum between consecutive samples. The step response is a step with a linear rising edge. The amplitude response shows a pass-band with gain of $6~dB$ from DC up to higher frequencies and a zero at Nyquist $z=-1$.}
    \label{fig:sys2_resp}
\end{figure}

Fig.~\ref{fig:block_diagram}-(b) depicts a FIR architecture implementing a two-sample moving sum. In the case of a zero at $z^{-1}=-1$, the transient responses shown in Fig.~\ref{fig:sys2_resp} demonstrates the effect of the two-units delay as the input impulse is broadened and the step's rising edge is linearly interpolated. The equivalent CT frequency response, exhibits an all-pass filtering behavior with a gain of $6~dB$, due to the two-sample summation. The gain roll-off near the Nyquist frequency arises because successive samples span different half‑cycles of the input sinusoid. The resulting progressive phase shift leads to increasing cancellation in the summation, culminating in a complete null at the Nyquist point where $z^{-1}=-1$.

\begin{figure}
    \centering
    \includegraphics[width=0.5\textwidth]{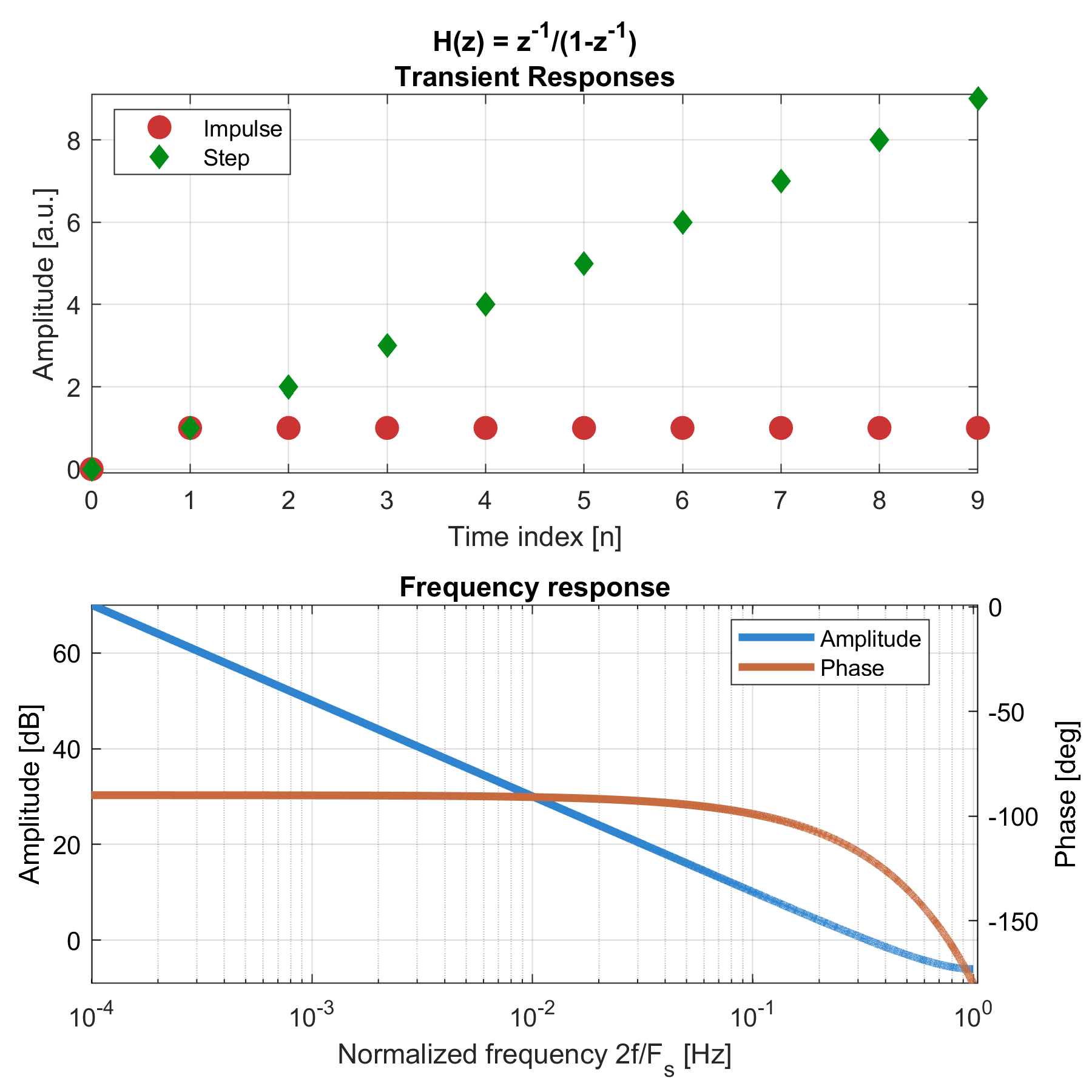}
    \caption{Transient (top) and frequency (bottom) response of the discrete accumulator. The impulse response is a step which is the integral of the impulse. Under convolution it produces accumulation with unitary weights of input samples. The step response is an unbounded linear ramp. The amplitude response shows a low-pass characteristic with a pole at DC $z=1$ and a minimum gain of $-6~dB$ at Nyquist $z=-1$.}
    \label{fig:sys3_resp}
\end{figure}

Fig.~\ref{fig:block_diagram}-(c) depicts an IIR architecture which with a unit pole reproduces an accumulator (ideal discrete integration). The transient responses in Fig.~\ref{fig:sys3_resp} illustrate the expected characteristic behavior by which an impulse is converted into a unit step and a step into a linear ramp. 
The equivalent CT frequency response reveals a pole responsible for infinite gain at DC. Whereas, near the Nyquist frequency the gain reaches a minimum of $-6~dB$ reflecting the progressive cancellation due to phase misalignment. 
The pole reflects the unbounded growth of the step response. Ideally a discrete integrator accumulates without limit in practice,  however, saturation constraints the operational range. Alternatively, free‑running accumulators can be employed within control loops, where they act as integrators whose unbounded drift is stabilized by negative feedback, effectively converting open‑loop integration into stable, closed‑loop integral sensing.

\begin{figure}
    \centering
    \includegraphics[width=0.5\textwidth]{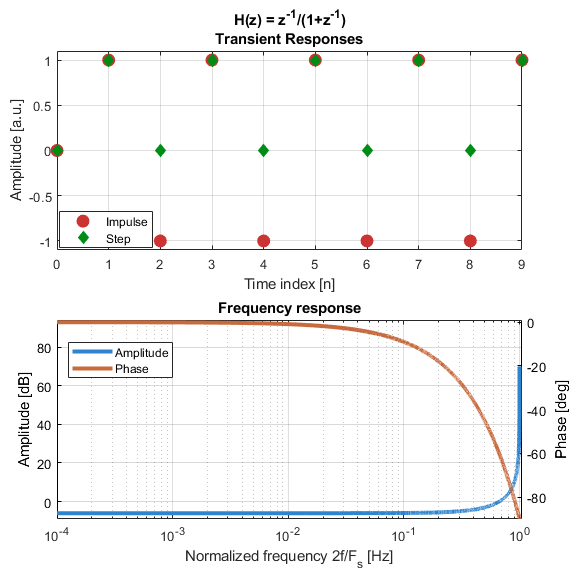}
    \caption{Transient (top) and frequency (bottom) response of the discrete oscillator. The impulse response is an oscillation at Nyquist which under convolution provides the sum of differences of pairs of consecutive samples. The step response is also an oscillation with smaller amplitude. The amplitude response shows a high-pass notch characteristic with a pole at Nyquist $z=-1$ and a minimum gain of $-6~dB$ away from Nyquist.}
    \label{fig:sys4_resp}
\end{figure}

Fig.~\ref{fig:block_diagram}-(d) depicts an IIR architecture implementing a DT oscillator. The transient responses with a unit negative pole shown in Fig.~\ref{fig:sys4_resp} presents a sustained oscillation at the Nyquist frequency.
The CT equivalent frequency response reveals an unstable pole at $z^{-1}=-1$, which produces infinite gain at that frequency while attenuating all others. Despite its intrinsic instability, this configuration is commonly employed as a clock‑frequency divider incorporating a sampled bi-stable element.

Another design method is to approximate an s‑domain transfer function, such as the first‑order low‑pass filter in Eq.~\ref{eq:low_pass_filter}, using the mapping in Eq.~\ref{eq:s_approx_z}. This yields:

\begin{equation}
    H_{\text{LPF}}(z) = \frac{T_s}{\tau + T_s} \cdot \frac{z^{-1}}{1 - \frac{\tau}{\tau + T_s} \cdot z^{-1}}
\end{equation}

Which can be written in a compact form highlighting the pole $z_p = \frac{\tau}{\tau + T_S}$: 
\begin{equation}
    H_{\text{LPF}}(z) = \frac{(1 - z_p) \cdot z^{-1}}{1 - z_p \cdot z^{-1}}
    \label{eq:lpf_discr_tf}
\end{equation}

\begin{figure}
    \centering
    \includegraphics[width=0.5\textwidth]{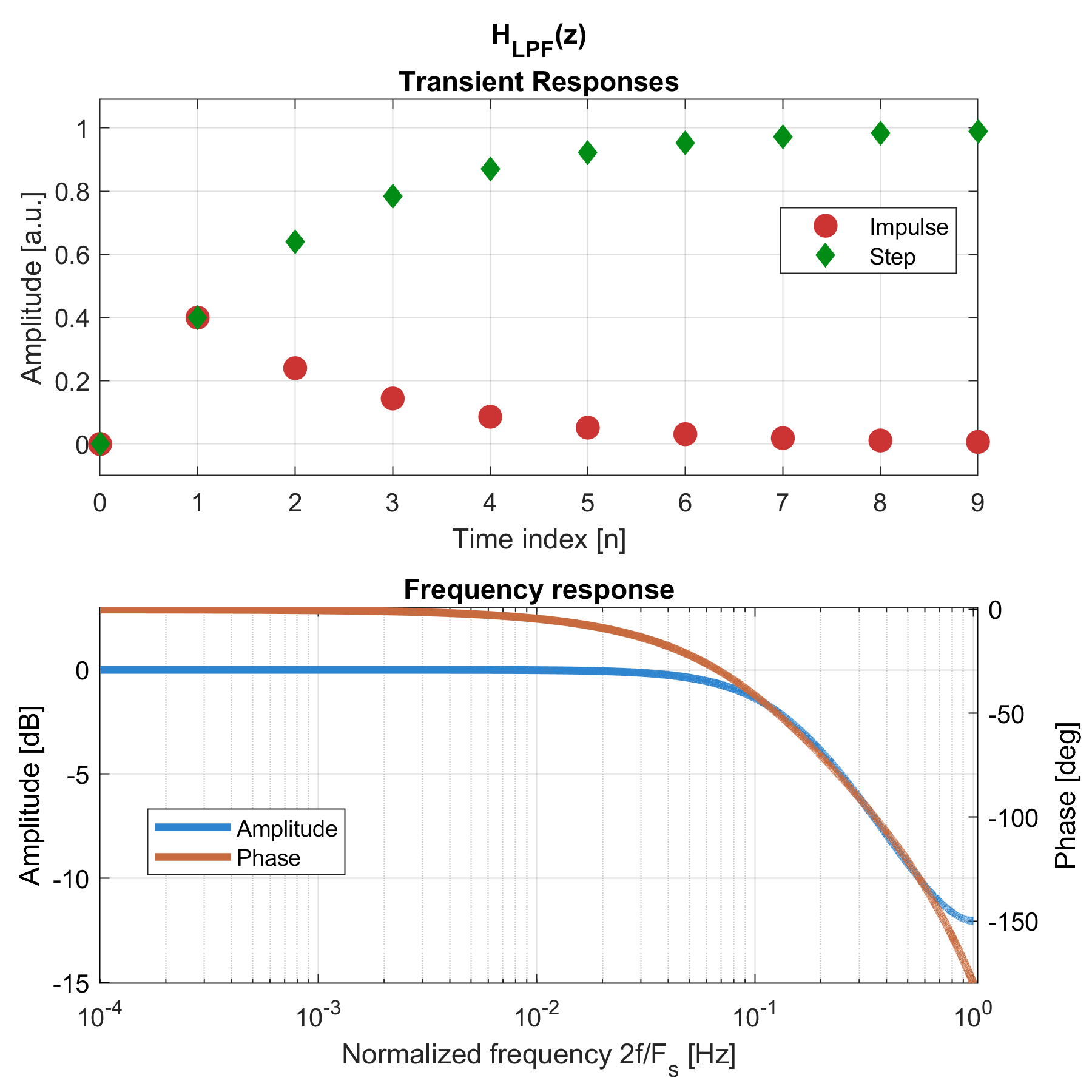}
    \caption{Transient (top) and frequency (bottom) of the discrete low-pass filter with $z_p=0.6$. The impulse response is a low-passed impulse with the characteristic exponential decay. The step response is a step with exponential settling. The amplitude response shows a low-pass characteristic with a pole at $z_p$ and a minimum gain at Nyquist $z=-1$.}
    \label{fig:LPF_resp}
\end{figure}

The corresponding block diagram can be drawn from the discrete accumulator, Fig.~\ref{fig:block_diagram}-(c), by incorporating the scaling factor associated with the pole $z_p$. The transient and frequency responses of Fig.~\ref{fig:LPF_resp} show the characteristic behavior of a leaky integration for frequencies well below the Nyquist point where the gain reaches its minimum $(1-z_p)/(1+z_p)$. This shows how the DT approximation only allows a limited attenuation of high-frequencies whose value is directly related to the pole responsible for the cut-off.

\begin{figure}
    \centering
    \includegraphics[width=0.5\textwidth]{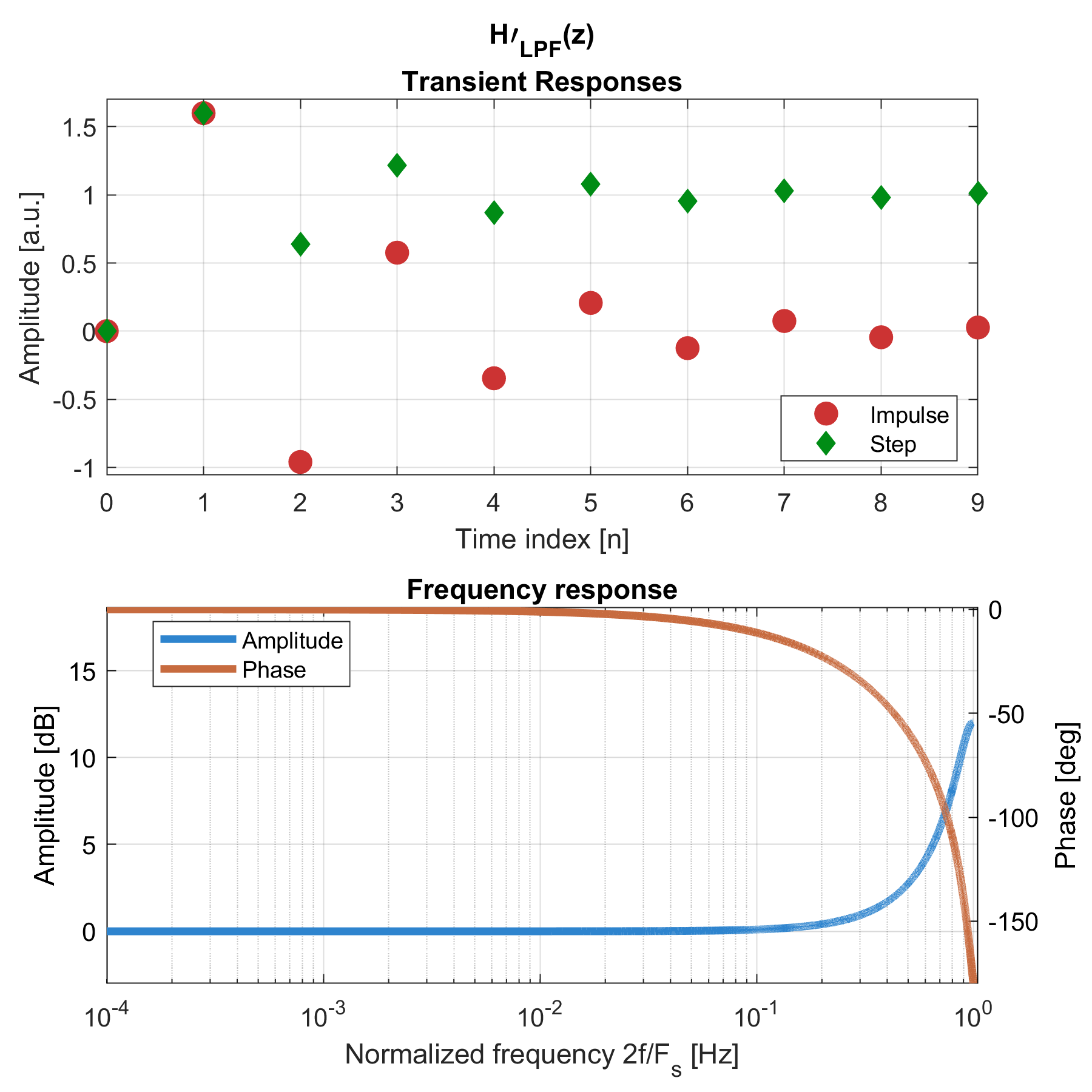}
    \caption{Transient (top) and frequency (bottom) of a discrete first-order transfer function with negative zero $z_p=-0.6$. The impulse response is an oscillation at Nyquist with an exponential decay envelope. The step response is an underdamped step with exponential settling. The amplitude response shows a gradual high-pass notch characteristic with a negative pole at $z_p$ a maximum gain at Nyquist $z=-1$.}
    \label{fig:LPF_inv_resp}
\end{figure}

Unlike CT, where the sign of the real part controls growth/decay, in DT both positive and negative real poles inside the unit circle are stable, but a negative pole induces alternating sign oscillations. The limit case $z_p=-e^{i\pi}$ corresponds to the discrete oscillator depicted inFig.~\ref{fig:block_diagram}-(d); where the sign alternation arises because the impulse response consists of integer powers of the pole, as illustrated in Fig.~\ref{eq:impls_resp_pole}. The pole's phase induces an oscillation at the Nyquist frequency which can be interpreted as a train of discrete doublets.
Inside the unit circle stability is insured by the damping action of the pole's modulus, which imposes a decaying exponential envelope on the oscillation. Alternatively, the impulse response can be interpreted as a single doublet (typically realized with a differentiating positive zero) subjected to leaky accumulation by an IIR integrator.    
The overall impulse response, reproduced in Fig.~\ref{fig:LPF_inv_resp}, when convolved with an input sequence results in a leaky accumulation of past differences.
Thus, a first-order system with a negative pole approximates a high‑pass filter by emulating a positive zero stabilized with a low‑pass pole; indeed, the resulting frequency response amplifies frequencies near the Nyquist point relative to lower ones, according to its transfer function:

\begin{equation}
H'_{\mathrm{LPF}}(z) = (1 + z_p) \cdot \frac{ z^{-1}}{1 + z_p \cdot z^{-1}}
\end{equation}

The corresponding block diagram can be obtained from the discrete oscillator of Fig.~\ref{fig:block_diagram}-(d) by adding a proper scaling factor in the feedback path  that moves the pole inside the unit circle.
The derivative characteristic is also observable in the step response where the amplification of the rising edge induces an overshoot typical of underdamped systems which are designed near their stability boundary; in the continuous case this condition corresponds to a reduced phase margin.

\begin{figure}
    \centering
    \includegraphics[width=0.5\textwidth]{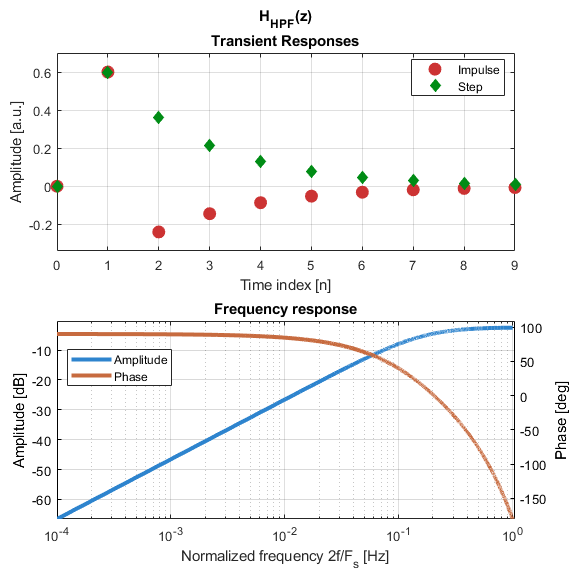}
    \caption{Transient (top) and frequency (bottom) of the discrete high-pass filter with $z_p=0.6$. The impulse response is a high-passed impulse with the characteristic exponential recovery. The step response is an impulse with exponential decay. The amplitude response shows a high-pass characteristic with a pole at $z_p$ and a maximum gain at Nyquist $z=-1$.}
    \label{fig:HPF_resp}
\end{figure}

Similarly the discrete first-order high-pass filter can be converted from Eq.~\ref{eq:high_pass_filter}, leading to: 

\begin{equation}
H_{\text{HPF}}(z) = z_P \cdot \frac{1 - z^{-1}}{1 - z_P \cdot z^{-1}} \cdot z^{-1}
\end{equation}
The corresponding block diagram can be obtained through the series connection of the differentiator, Fig.~\ref{fig:block_diagram}-(a), and the accumulator, Fig.~\ref{fig:block_diagram}-(c), with a proper scaling in the feedback to position the pole. This effectively adds a zero at DC to the low-pass filter of Eq.~\ref{eq:lpf_discr_tf}.
The transient and frequency responses in Fig.~\ref{fig:HPF_resp} exhibit a well‑behaved high‑pass characteristic across the entire Nyquist band. In contrast, the corresponding low‑pass filter shows increasing deviation as the frequency approaches the Nyquist point, where aliasing and sampling effects become pronounced

The digital implementation of DT filters offers distinct advantages, most notably the ability to adaptively interconnect subsections in real-time to realize a desired transfer function and filter-order. This enables dynamic optimization of performance metrics such as power consumption, gain, and rejection.
Additional flexibility is afforded by multi-rate operation, whereby the equivalent $s$-domain characteristic can be tuned by adjusting the operating frequency of the digital hardware. This is possible because the poles and zeros depend on the sampling frequency through the mapping given in Eq.~\eqref{eq:pole_zero_map}, which can be approximated as: 
\begin{equation}
    z_{p,z} = 1 + s_{p,z} T_s + \frac{(s_{p,z} T_s)^2}{2!} + \frac{(s_{p,z} T_s)^3}{3!} + \cdots,
\end{equation}

\section{Conclusions}
The apparent divide between CT and DT signal processing often obscures the fact that both domains describe the same underlying dynamical behaviors subject to different operational constraints. By examining first-order systems across the Laplace and Z domains, this work has shown that concepts such as stability, causality, memory, and frequency selectivity retain their meaning even as their mathematical representations change. The limitations encountered in physical implementations—finite bandwidth, saturation, delay, and noise—are not artifacts of approximation but intrinsic features that shape realizable systems. DT filters do not merely approximate their CT counterparts; they reinterpret them through sampling and accumulation. Recognizing this continuity provides a more robust intuition for system design, particularly in mixed-signal and digital implementations where naive translations between domains often fail. While the analysis here has focused on first-order structures, the same principles extend naturally to higher-order, multirate, and noise-shaped systems, suggesting that conceptual clarity remains central to practical signal processing.


\end{document}